\documentstyle[aaspp4]{article}

\begin{document}

\title{Thermal conduction in accretion disk coronae}
\author{Andrzej Macio\l ek--Nied\'zwiecki\altaffilmark{1}, 
Julian H. Krolik\altaffilmark{2} and Andrzej A. Zdziarski\altaffilmark{1}}

\altaffiltext{1}{N. Copernicus Astronomical Center, Bartycka 18, 00-716 
                Warsaw, Poland}

\altaffiltext{2}{Department of Physics and Astronomy, Johns Hopkins University,
                 Baltimore, MD 21218, USA}

\begin{abstract}
We study the effects of thermal conduction in a hot, active corona above an
accretion disk. We assume that all of the dissipative heating takes place in
the corona. We find that the importance of conduction decreases with increases
in the local dissipative compactness of the corona, $\ell_{\rm diss,loc}$, and
increases with increasing abundance of electron-positron pairs. For $\ell_{\rm
diss,loc}<1$, a significant fraction of the energy released in the corona may
be carried away by the conductive flux, leading to formation of a relatively
hot transition layer below the base of the corona.  Comptonization of disk
radiation in such a layer may account for the presence of soft X-ray excesses
in the spectra emitted by disk-corona systems.
\end{abstract}


\section{INTRODUCTION}

An accretion disk with a dissipative, thermal corona has recently become a
popular model for interpreting the UV/X/$\gamma$-ray spectra from Seyfert
galaxies and Galactic black hole candidates (see, e.g., Svensson 1996). 
Recent studies
have focused on the correct description of the spectrum from such a system
(e.g., Stern et al.\ 1995, Poutanen \& Svensson 1996, Sincell \& Krolik 1997)
whereas relatively little
effort has been taken to achieve a physically self-consistent internal
structure of the corona. In particular, most of the existing models of
disk-corona systems involve two simplifying assumptions: isothermal structure
of the corona, and purely radiative energy exchange between the corona and the
disk.  We relax the above assumptions by considering effects of thermal
conduction, which will obviously occur in a hot corona situated above the
surface of a cold disk. Such a treatment allows us to determine
self-consistently the electron density and temperature distributions in the
corona.

We consider two limiting cases: a corona consisting purely of electron-positron
pairs, and a corona without pairs (a pure electron-proton plasma). The
calculations of thermal pair equilibria (e.g., Stern et al.\ 1995) show that
both cases ($e^\pm$ and $ep$ dominated coronae) may be expected in accreting
black hole systems, depending on both the compactness (defined below) and the
temperature.  We assume that $ep$ coronae satisfy the condition of hydrostatic
equilibrium.  For $e^\pm$ coronae, the hydrostatic model is inconsistent  and
large-scale coronal expansion is expected, as we discuss in section 2.5.
Leaving the investigation of the exact structure of pair coronae for future
studies, we consider here a constant pressure structure to see what effect
conduction may have in the presence of pairs.
 
\section{PHYSICAL PROCESSES IN A CORONA}

\subsection{Notation and basic assumptions}
We assume that the corona has a slab geometry with a scale height $h$ and a
surface area $A$. We also assume a sandwich geometry (coronae located on both
sides of the disk) so that we can consider only half of such a plane
symmetrical system. The plasma in the corona is assumed to be optically thin
($\tau_{\rm tot}\lesssim 1$, where $\tau_{\rm tot}= \sigma_T \int_0^h n_e(z) dz$
is the
total optical depth, $n_e$ is the electron density, and $z$ is the distance
from the corona bottom) and to have a thermal distribution of electron energies
(with the temperature varying across the corona). These assumptions are
supported by observations of Seyfert 1 galaxies (e.g., Zdziarski et al.\ 1995;
Gondek et al.\ 1996) and Galactic black hole candidates (e.g., Gierli\'nski et
al. 1997), whose spectra are well fitted by optically thin thermal
Comptonization.

We measure the electron density and the temperature of the plasma by 
the following dimensionless parameters,
\begin{equation}
\Theta \equiv {kT \over m_{\rm e}c^2},
\end{equation} 
\begin{equation}
\tau_x \equiv n_e(x)\sigma_Th, 
\end{equation}
where $x \equiv z/h$. The latter parameter is the differential Thomson depth
at $x$ in the interval $dx$, i.e. $\tau_x=d\tau / dx$, where $d\tau=n_e(z)
\sigma_T dz$. 

Following Haardt \& Maraschi (1991, 1993), we assume that most of the accretion
energy is dissipated in the corona. The dissipated power is lost due to thermal
conduction and radiation losses. Half of the high energy radiation from the
corona irradiates the cold disk and the other half escapes directly from the
source (we neglect for simplicity the anisotropy of the scattering process in
the corona and reflection of hard photons from the cold matter).  The coronal
radiation intercepted by the disk is reprocessed and reemitted at much lower
energies, which provides soft seed photons for Comptonization in the corona.
The dissipated power, $L_{\rm diss}$, and both the corona and disk radiation
luminosities, $L_{\rm h}$ and $L_{\rm s}$, respectively, are characterized by
compactness parameters:
\begin{equation}
\ell_{\rm diss} \equiv {L_{\rm diss} \over h} {\sigma_T \over m_e c^3},~~~
\ell_{\rm h} \equiv {L_{\rm h} \over h} {\sigma_T \over m_e c^3}, ~~~
\ell_{\rm s} \equiv {L_{\rm s} \over h} {\sigma_T \over m_e c^3}.
\end{equation}
[$L_{\rm h}$ and $L_{\rm s}$ correspond to the total power of radiation emitted
by one side of the corona and the disk, while the observed luminosities of
those layers are described by $L_{\rm h}/2$ and $\sim L_{\rm s} \exp
(-\tau_{\rm tot})$, respectively.] Our main parameter is, however, the local
dissipative compactness parameter (introduced by  Bj\"ornsson \& Svensson
1992),
\begin{equation}
\ell_{\rm diss,loc} \equiv {L_{\rm diss,loc} \over h} {\sigma_T \over m_e c^3},
\end{equation}
where $L_{\rm diss,loc}=L_{\rm diss} h^3 /(Ah)$, and thus
\begin{equation}
\ell_{\rm diss,loc}=\ell_{\rm diss} {h^2 \over A}.
\end{equation} 
This parameter is crucial for determining the efficiency of radiative processes
in the corona, since it involves the luminosity in a cubic volume with the size
$h$, which is the characteristic local volume in the corona.
Analogously, we define the local radiative compactnesses,
\begin{equation}
\ell_{\rm h,loc}=\ell_{\rm h} {h^2 \over A},~~~
\ell_{\rm s,loc}=\ell_{\rm s} {h^2 \over A}.
\end{equation}

We emphasize the importance of the difference between the local compactness
and global compactness, $\ell_{\rm glob} \equiv L \sigma_T /(R m_e c^3)$,
where $R$ is the distance from the central source, in the corona geometry (see
also Gondek et al. 1996).  It is the local compactness which is the most
relevant for parameterizing the character of the coronal thermal and pair
balance, which fact is clear from the form of our equations
(\ref{e1}--\ref{e3}) below.  It is equally true for other problems, for
example, for the issue of whether the electrons in the corona can achieve a
thermal distribution function (cf. Ghisellini et al. 1993).   The electron
distribution function is determined by a competition between the
electron-electron relaxation timescale (determined by the electron density and
mean energy) and the electron cooling timescale (which depends on the
radiation energy density).  Their ratio is then a function of the optical
depth and the {\it local} compactness, not the {\it global} compactness.
Since $\ell_{\rm loc}$ is always smaller than $\ell_{\rm glob}$, and sometimes
very much smaller, this can be an important distinction.  The cooling time
becomes relatively longer as $\ell$ falls, so using $\ell_{\rm glob}$ instead
of $\ell_{\rm loc}$ can lead to a spurious difficulty in achieving an
equilibrium distribution function.

The importance of the conduction process will be measured by the fraction of 
the dissipated power transported out of the corona by the conductive flux,
\begin{equation}
\epsilon_q={|q_0| A \over L_{\rm diss}},
\label{epsq}
\end{equation}
where $q_0$ is the heat flux through the corona bottom.

Taking into account both the radiative energy exchange and the conductive 
transport of thermal energy, we can relate the corona and disk luminosities 
to the dissipated power,
\begin{equation}
L_{\rm h}=L_{\rm diss}(1-\epsilon_q),~~~~~~
L_{\rm s}=0.5L_{\rm h}+\epsilon_q L_{\rm diss},
\label{lhs}
\end{equation}
(we neglected here the initial energy of soft photons scattered in the corona 
as they contribute negligibly to the corona luminosity). Then
\begin{equation}
{\ell_{\rm s} \over \ell_{\rm diss}}=0.5\left ( 1 + \epsilon_q \right ),~~~~~~
{\ell_{\rm h} \over \ell_{\rm diss}}=1 - \epsilon_q.
\label{qa}
\end{equation}

\subsection{Heating}
We assume that all electrons in the corona are heated at a constant rate,
$\gamma$. Then the heating rate per unit volume is,
\begin{equation}
H=\gamma n_e(z),
\label{h2}
\end{equation}
which corresponds to a local dissipative compactness,
\begin{equation}
\ell_{\rm diss,loc}=\Gamma \tau_{\rm tot}, 
\label{ee}
\end{equation}
where 
\begin{equation}
\Gamma={\gamma \over m_e c^2} {h \over c}.
\end{equation}

\subsection{Radiative processes}

Radiative cooling in the corona is due to unsaturated Comptonization and 
bremsstrahlung. The Compton cooling rate is 
\begin{equation}
C_{\rm Compt}=\max(4\Theta,16\Theta^2)n_{\rm e}c \sigma_{\rm T}
U_{\rm rad},
\end{equation}
where $U_{\rm rad}$ is the radiation energy density, $n_{\rm e}$ is the
electron density, and the expressions in parentheses give the average relative
photon energy gain per scattering in the nonrelativistic ($4\Theta$) and
relativistic ($16\Theta^2$) cases, respectively (for $0.2<\Theta<0.3$, we use
$80\Theta^3-32\Theta^2+7.2\Theta$ in order to achieve a smooth transition
between the two regimes). We take into account cooling of electrons on both
soft photons from the disk and photons already scattered in the corona.  As
photons scattered many times in the corona may enter the Klein-Nishina regime,
we include only photons satisfying the condition
\begin{equation}
\varepsilon < \min(\Theta^{-1},1)
\end{equation}
where $\varepsilon \equiv h \nu /m_{\rm e}c^2$. Then,
\begin{equation}
U_{\rm rad}={1 \over c}{L_{\rm s}+\eta L_{\rm h} \over A},
\end{equation}
where the fraction of the energy density that scatters in the Thomson regime is
\begin{equation}
\eta ={\int_{\varepsilon_0}^{\min(1/\Theta,1)} F(\varepsilon) d \varepsilon 
\over \int_{\varepsilon_0}^\infty F(\varepsilon) d \varepsilon},
\label{r}
\end{equation}
$F(\varepsilon)$ is the energy spectrum of the hard radiation from the corona
[we assume a power-law spectrum with the energy index given by eqs.\ (1) and
(2) in Zdziarski et al. (1994) and an exponential cut-off at $E_c=1.6kT$], and
$\varepsilon_0$ is the soft photons' energy. The bremsstrahlung cooling rate is
\begin{equation}
C_{\rm brem}=n_e^2 m_ec^3\sigma_T l_{\rm brem}(\Theta), 
\end{equation}
where 

\begin{equation}
l_{\rm brem}(\Theta)=\cases{
4 \sqrt{2} \alpha_f \pi^{-3/2} \left(\Theta^{1/2}
+1.781\Theta^{1.84} \right),               & $\Theta \le 1$, \cr
4.5 \alpha_f  \pi^{-1} \left[ \ln(1.12 \Theta+0.42)+1.5 \right], 
                                           & $\Theta \ge 1$,  \cr}
\label{lb}
\end{equation}
and $\alpha_f$ is the fine-structure constant (Svensson 1982). 

\subsection{Thermal conduction}

The classical heat flux is expressed by
\begin{equation}
q_{\rm class}=-\kappa {dT \over dz},
\end{equation}
where $\kappa$ is the thermal conductivity. In a fully ionized hydrogen 
plasma,
\begin{equation}
\kappa=\phi_c \chi T^{5 \over 2},
\end{equation}
where the parameter $\phi_c$ allows for a reduction of the heat flux by 
magnetic fields, turbulence etc.\ (and is taken here to be $\phi_c=1$) and
\begin{equation}
\chi= 18 \left ( {2 \over \pi} \right )^{3/2} {k^{7/2} \over m_e^{1/2}e^4 
\ln \Lambda}
\end{equation} 
(Spitzer 1962), where
\begin{equation}
\ln \Lambda=29.7+\ln \left[ (T_e/10^6{\rm K}) n_e^{-1} \right ]
\end{equation}
is the Coulomb logarithm. As $\ln \Lambda$ depends logarithmically on electron
density and temperature, we assume a constant value of $\ln \Lambda=30.7$,
which corresponds to $\chi=6 \times 10^{-7}{\rm erg~s}^{-1}{\rm cm}^{-1} {\rm
K}^{-7/2}$. The classical expression for the heat flux is based on the
assumption that the electron mean free path is short compared to the
temperature height scale, $T/ | \nabla T | $. In the limit $| \nabla T |
\rightarrow \infty$ the heat flux is said to be ``saturated", and Cowie \&
McKee (1977) suggested the form
\begin{equation}
| q_{\rm sat} | =5 \phi_s \rho c_s^3, 
\label{qsat}
\end{equation}
where $c_s=(kT/\mu)^{1/2}$ is the isothermal sound speed, $\mu$ is the mean
mass per particle ($\mu \approx m_p/2$ for $ep$ plasmas and $m_e$ for $e^\pm$
plasmas), $\rho$ is the mass density and $\phi_s$ parametrizes the uncertainty
of the appropriate value of $q_{\rm sat}$. Both experimental and theoretical
results constrain $\phi_s$ to the range $\phi_s=0.2$--$1.1$ with some
preference for the value of $\phi_s=0.3$ (Balbus \& McKee 1982; Giuliani
1984). We use here $\phi_s=0.3$ but test the sensitivity of the results to the
value of $\phi_s$. The saturated heat flux given by eq.\ (\ref{qsat})
corresponds to advection of the plasma's thermal energy at a rate comparable
to the sound speed. In order to smoothly implement the transition from
classical diffusive to saturated trasport, Balbus \& McKee (1982) defined the
effective heat flux as the harmonic mean of $q_{\rm sat}$ and $q_{\rm class}$,
\begin{equation}
q=-{\kappa \over 1+\sigma} {d T\over dz},
\end{equation}
where
\begin{equation} 
\sigma= \left| {q_{\rm class} \over q_{\rm sat}} \right|. 
\end{equation}
The equation of heat transfer in the vertical direction,
\begin{equation}
{d q \over dz}=H-C_{\rm Compt}-C_{\rm brem},
\label{cond}
\end{equation}
may then be written in a dimensionless form,
\begin{equation}
{d \over dx} \left ( {a \Theta^{5/2} \over 1+b\Theta|\Theta'|}\Theta' \right )
=f(\Theta ,\tau_x),
\label{eq1}
\end{equation}
where
\begin{equation}
f(\Theta ,\tau_x)=\ell_{\rm diss,loc} \tau_x {\ell_{\rm s} +\eta \ell_{\rm h} 
\over \ell_{\rm diss}} \max(4 \Theta,16 \Theta^2)-\Gamma \tau_x 
+ l_{\rm brem}(\Theta) \tau_x^2, 
\label{f}
\end{equation}
\begin{equation}
a \equiv \phi_c \chi \left ( {m_e c^2 \over k} \right )^{7/2} {\sigma_T \over 
m_e c^3}, 
\end{equation}
\begin{equation}
b \equiv {1 \over \tau_x}{a \over 5 \phi_s \delta}  \left ( {\mu \over m_e}
\right ) ^{1/2},
\end{equation}
\begin{equation}
\Theta' \equiv {d \Theta \over dx},
\end{equation}
and $\delta=$1 and 2 for electron-positron and electron-proton plasmas, 
respectively.

\subsection{Hydrostatic equilibrium}

For an electron-proton corona, the condition $h \ll R$ is satisfied and the 
following approximate form of the hydrostatic equilibrium equation may be used,
\begin{equation}
{dP \over dz}=-{GM \over R^3}(z+z_d) \rho(z), 
\label{hydro}
\end{equation}
where $P$ is the pressure in the corona and $z_d$ is the distance of the corona
bottom from the disk midplane. The pressure is the sum of radiation and gas 
pressures. From
\begin{equation}
{P_{\rm rad} \over P_{\rm gas}} = \ell_{\rm diss,loc} {\ell_s+\ell_h
\over \ell_{\rm diss}} {1 \over 2 \Theta \tau},
\end{equation}
$\Theta \tau \approx 0.1 (\ell_h/\ell_s)^{1/4}$ (Pietrini \& Krolik 1995), and
eq.\ (\ref{qa}) (with $\epsilon_q < 0.5$, as found in section 4), we find that
the gas pressure dominates for $\ell_{\rm diss,loc} < 0.15$. At higher
compactnesses thermal conduction becomes unimportant (section 4), so we
neglect radiation pressure. Then equation (\ref{hydro}) with $P=2n_ekT$ gives
\begin{equation}
{d \tau_x \over dx}= -{\tau_x \over \Theta} \left [ \Theta'+{1 \over 4r}
{m_p \over m_e} \left ( {h \over R} \right ) ^2 (x+x_d)  \right],
\label{eq2}
\end{equation}
where $x_d=z_d/h$, $r=R/R_{\rm Sch}$ and $R_{\rm Sch} \equiv 2GM/c^2$. 
For an isothermal corona the above equation gives,
\begin{equation}
(h/R)^2=8r{m_e \over m_p}\Theta_c,
\label{hr}
\end{equation}
where $\Theta_c$ is the (constant) temperature of the corona.  In section 4 we
find that thermal conduction significantly affects the corona structure near
its base in a way that may increase the corona total optical depth. However,
the temperature at higher altitudes is roughly constant and the density
profile does not differ from that found in isothermal approximation.
Therefore, we use the scale hight determined by eq.\ (\ref{hr}), approximating
$\Theta_c$ by the temperature at the top of the corona, $\Theta_{\rm top}$
(see section 2.7).

The distance, $z_d$, is the sum of the disk height, $h_{\rm disk}$, and the
height of the transition layer affected by thermal conduction, $h_{\rm tl}$.
Using the results of Svensson \& Zdziarski (1994), we find that the ratio of
corona height to disk height is $h/h_{\rm disk} \gtrsim 100$.  The value of
$h_{\rm tl}$ is determined by the solution of the conduction problem and (as
we find in section 4) typically $h_{\rm tl} \gtrsim 0.1h$ when conduction is
important.  In this case, the value of $z_d$ is dominated by the height of the
transition layer. The exact value can be obtained with additional parameters
of the model (accretion rate, viscosity parameter etc.) necessary for the
calculation of $h_{\rm disk}$.

For electron-positron coronae, the gravitational force acting on the corona is
too small to keep it bound because the escape speed is $\sim c r^{-1/2}$,
whereas the sound speed is $\sim c \min(1/\sqrt{3}, \Theta^{1/2})$ and Compton
equilibrium generically gives temperatures $\Theta$ not much less than unity.
Consequently, in a pure pair plasma $h \sim R$ and strong outflows can be
expected unless the plasma is confined by magnetic fields.

In order to check the importance of thermal conduction in pair coronae, we
consider here a zone of $e^\pm$ plasma at low altitudes above the disk (as it
is unlikely that the heating mechanism operates at large distances from the
disk) close to hydrostatic equilibrium so that the corona exists.  As the
pressure gradient induced by the pair gravity is negligible, we expect that
such a zone is almost isobaric and so we will describe that zone by eq.\
(\ref{eq2}) with the null gravity term (the second term in brackets).
 
\subsection{Transition layer}
As we find in section 4 below, the bottom of the corona is much hotter than
the disk. Thus a transition layer between the disk and the corona must exist
over which the temperature decreases from the coronal to disk value. Decrease
of the temperature in the transition layer is accompanied by a decrease of
conductive heat flux, as the energy transported from the disk is gradually
radiated away. At some point the conductive flux vanishes and the temperature
levels off at a value determined by the balance between radiative heating and
cooling. We call this point the base of the transition layer. Below the base
the temperature is still higher than the ``normal" disk temperature,
due to Compton
heating by the coronal radiation. We do not investigate properties of that
Compton heated layer as atomic processes (which are neglected in this paper)
may become important there (a detailed analysis of such regions has recently
been done by R\'o\.za\'nska \& Czerny 1996 and Sincell \& Krolik 1997).

The structure of the transition layer affected by thermal conduction  may be
found in the framework of our model with $H=0$. We take into account Compton
heating of electrons in the transition layer. Then the right-hand side of the
heat transfer equation (\ref{eq1}) takes the form,
\begin{eqnarray}
f(\Theta ,\tau_x) & = & \ell_{\rm diss,loc}\tau_x \biggl\{ 0.5 \eta \exp
(-\int_x^0 \tau_x dx) \ell_{\rm h} \ell_{\rm diss}^{-1} \left [ 
\max(4\Theta,16\Theta^2)
-\langle \varepsilon \rangle \right ] \nonumber \\
& & +~~~\ell_s  \ell_{\rm diss}^{-1} \left [ \max(4\Theta,16\Theta^2)-
\langle \varepsilon_0 \rangle \right ] \biggr\} + l_{\rm brem}(\Theta)
\tau_x^2, 
\label{f1} 
\end{eqnarray}
where $\eta$ is given by eq.\ (\ref{r}) and
\begin{equation}
\langle \varepsilon \rangle = {\int F \varepsilon d \varepsilon
\over \int F d \varepsilon}.
\label{vareps}
\end{equation}
The integrals over the energy spectrum of the radiation emitted from the
corona have upper limits equal of $\min(\Theta^{-1},1)$, and $\langle
\varepsilon_0 \rangle = 5 \times 10^{-5}$ is assumed for soft radiation from
the disk. (For Galactic black hole candidates, 
$\langle \varepsilon_0 \rangle = 10^{-3}$ would be a
more appropriate value, but the results depend weakly on this parameter.)

While for $ep$ coronae the transition layer will obviously consist of an
electron-proton plasma,  the composition of the transition layer for $e^\pm$
coronae is rather uncertain and depends on the mechanism of formation of the
transition layer. If it is formed by the heating of the outer parts of the
accretion disk, we can expect that it is composed mostly of $ep$ with some
addition of pairs inflowing from the corona. On the other hand, the transition
layer may be formed together with the active region and then it would be
dominated by electron-positron plasma. We will examine here the latter
possibility (by assuming that both the corona and the transition layer consist
only of $e^\pm$ plasma) to estimate the maximum effect of thermal conduction,
which is achieved for a pure pair plasma (see section 4).

\section{CALCULATIONAL PROCEDURE}

\subsection{Differential Equations}

Eqs.\ (\ref{eq1}) and (\ref{eq2}) give three first-order differential 
equations, 
\begin{equation}
{d \Theta \over dx}=\Theta',
\label{e1}
\end{equation}
\begin{equation}
{d \Theta' \over dx}=f(\Theta ,\tau_x){(1+b \Theta \Theta')^2 \over a 
\Theta^{2.5}}+b{\Theta'}^2 g (x+x_d)  -0.5b{\Theta'}^3-2.5{{\Theta'}^2 
\over \Theta},
\label{e2}
\end{equation}
\begin{equation}
{d \tau_x \over dx}= -{\tau_x \over \Theta} \left [ \Theta'+ g (x+x_d)  
\right],
\label{e3}
\end{equation}
where $f(\Theta, \tau_x)$ is given by eqs.\ (\ref{f}) and (\ref{f1}) for the 
corona and the transition layer, respectively, and
\begin{equation}
g =\cases{ 2\Theta_c, & {\rm for proton-electron plasma}, \cr
           0,         & {\rm for electron-positron plasma}. \cr }
\end{equation}
The null value of $g$ for $e^\pm$ coronae results from the isobaric
approximation, whereas $g \equiv (h/R)^2 m_p/(4rm_e)$ for $ep$ coronae.  We
solve the above equations for $\Theta(x)$, $\Theta'(x)$ and $\tau_x$.

\subsection{Parameters and Boundary Conditions}
 
We are solving three first-order differential equations which depend on two
dimensionless parameters, $\tau_{\rm tot}$ and $\ell_{\rm diss,loc}$.  Thus,
in addition to choosing the parameters, we must also choose three boundary
conditions, $\Theta_{\rm top}$, $\Theta^{\prime}_{\rm top}$, and $\tau_x$ at the
top of the corona, which we call $\tau_{\rm top}$.

The physically appropriate choice for the temperature gradient is obvious.
There should be no heat flux at the top of the corona, so
\begin{equation}
\left. {d \Theta \over dx} \right |_{\rm top}=0. 
\label{b1}
\end{equation}

The other boundary conditions are subtler to determine.  In the absence of
conduction, the temperature throughout the corona would be determined by
$\ell_{\rm diss,loc}$ and $\tau_{\rm tot}$.  This is the calculation performed
in numerous other discussions of accretion disk coronae.  However, heat
conduction removes some heat from the corona, decreasing its temperature, and
we do not know how much this affects the top of the corona before doing the
calculation.   We implicitly fix $\Theta_{\rm top}$ by setting the temperature
at the bottom of the transition layer to be effectively zero, or at least far
below the coronal temperature.  The value used to initiate the integration is
then a parameter to be iterated on until self-consistency (see the Appendix
for details).

The third boundary condition, $\tau_{\rm top}$, may be thought of as the
density at the top of the corona.  Because we have normalized our unit of
distance to the scale height, it is implicitly fixed by $\tau_{\rm tot}$.  In
the case of an $ep$ plasma, the density profile of the corona is approximately
Gaussian (it would be exactly Gaussian if the temperature were exactly
independent of height), so formally the corona extends to infinity.  However,
we must choose some finite number of scale lengths at which to bound our
calculation.  We do so by setting the differential Thomson
depth at the top $\tau_{\rm top}
= 0.0001$.  If the corona were exactly isothermal, this would be achieved at
$x_{\rm top} = {\ln^{1/2} (\tau_{\rm tot}/\tau_{\rm top})}$, so by starting the
integration at $x_{\rm top}$ with $\tau_{\rm top} = 0.0001$, all would be
self-consistent.  To account for the deviations from isothermality induced by
conduction, we iterate on $x_{\rm top}$, keeping $\tau_{\rm top}$ fixed at
0.0001.  In the case of a pair plasma, our constant pressure assumption
translates to a density profile which approaches closer and closer to a
step-function in the limit as the temperature profile approaches isothermal.
We therefore start the calculation at $x=1$ and assume an initial value of the
coronal pressure.  Then we adjust the value of the pressure to obtain the
assumed $\tau_{\rm tot}$.

The optical depth, $\tau_{\rm tot}$, and the local compactness, $\ell_{\rm
diss,loc}$, of the corona are the main parameters of the model.  $x_d$ is also
a free parameter, but its value can be determined for a specific disk model, as
discussed in section 2.5.  Increase of $x_d$ results in a slight increase of
$\epsilon_q$, due to the change of the density profile in the corona. The
solution of the transition layer structure gives an obvious constraint on the
value of $x_d$, which must exceed the height of the transition layer. We assume
an initial value of $\epsilon_q$, which determines  $\ell_s/\ell_{\rm diss}$
and $\ell_h/\ell_{\rm diss}$ from eq.\ (\ref{qa}). The value of $\Gamma$
follows from eq.\ (\ref{ee}).

Summarizing, our model has three free parameters, $\tau_{\rm tot}$, $\ell_{\rm
diss,loc}$ and $x_d$. In calculations, we first adjust $\Theta_{\rm top}$ and
$x_{\rm top}$ (or $P_{\rm gas}$ for $e^\pm$ coronae) as described above.  Then
the value of $\epsilon_q$ from the solution with adjusted $\Theta_{\rm top}$
and $x_{\rm top}$ is used in the next iteration.  The procedure ends when the
calculated value of $\epsilon_q$ converges.  As discussed in Appendix A, once
the three free parameters are specified, there is a (narrow) allowed range
$(\Theta_{\rm top}, \Theta_{\rm top}+\Delta \Theta)$ (and a corresponding
range of $x_{\rm top}$) within which acceptable solutions exist.  The members
of this solution family are distinguished by the pressure at the bottom
of the transition zone.

For $ep$ coronae we computed solutions for $0.01 \leq \ell_{\rm diss,loc} \leq
0.4$, while in the pair case we found solutions for $0.01 \leq \ell_{\rm
diss,loc} \leq 10$.  The lower limit on $\ell_{\rm diss,loc}$ was suggested by
both the observational results [see, e.g., table 1 in Done \& Fabian (1989)
noting that the local compactness is roughly an order of magnitude lower than
the global compactness given there] and theoretical expectations [for slab
coronae $\ell_{\rm diss,loc} < 0.01$ corresponds to $L<10^{-3} L_{\rm Edd}$
(see section 5) which is unlikely in AGNs].  Note that for $\ell_{\rm
diss,loc} < 0.1$ the corona may still be pair dominated (see, e.g., figure 1
in Stern et al. 1995).  The upper limits are set by the character of our
solutions: at higher $\ell_{\rm diss,loc}$ conduction becomes negligible (see
below).  For both $e^\pm$ and $ep$ coronae, the optical depth range over which
we computed solutions was $0.06< \tau_{\rm tot} <0.3$.

\section{RESULTS}

We find that heat flux saturation puts an important constraint on the
conduction problem in that it establishes a limit  on the amount of thermal
energy that can be transported out of the corona. Namely, the conductive heat
flux at the bottom cannot exceed the saturated heat flux,
\begin{equation}
q_{\rm sat}=5 \phi_s \delta \left( {m_e \over \mu} \right )^{1/2}
{m_e c^3 \over \sigma_T h} \tau \Theta^{3/2}.
\end{equation}
As this limit is independent of the total amount of released power, the
relative importance of the thermal conduction process increases with
decreasing compactness. The maximum fraction of the released power that can
leave the corona in the form of conductive flux is,
\begin{equation}
{q_{\rm sat} A \over L_{\rm diss}}={5 \phi_s \delta (m_e/\mu)^{1/2} \tau
\Theta^{3/2} \over \ell_{\rm diss,loc}}.
\label{ratio}
\end{equation}
As discussed below, the temperature cannot differ strongly from the Compton 
equilibrium value, which can be estimated from $\Theta \tau \simeq 0.1 
(l_h/l_s)^{1/4}$ (Pietrini and Krolik 1995). Then $\ell_h/\ell_s \leq 2$ 
[eq.\ (\ref{qa})] implies,
\begin{equation}
{q_{\rm sat} A \over L_{\rm diss}} \leq 0.2 {\delta (m_e/\mu)^{1/2} 
\Theta^{1/2} \over \ell_{\rm diss,loc}},
\label{ratio1}
\end{equation} 
where we assumed $\phi_s=0.3$. As $\Theta \lesssim 1$ in both Seyfert galaxies 
and black hole binaries (e.g., Zdziarski et al.\ 1997), we find that
conduction has a rather minor effect (transporting less than 5 per cent of the
released power) for $\ell_{\rm diss,loc} \gtrsim 1$ and $\ell_{\rm diss,loc}
\gtrsim 0.15$ for electron-positron and electron-proton plasmas, respectively.

Figures 1 and 2 show the dependence of $\epsilon_q$ on $\ell_{\rm diss,loc}$.
The relation suggested by eq.\ (\ref{ratio1}), $\epsilon_q \propto
\ell_{\rm diss,loc}^{-1}$, appears to be valid for $\ell_{\rm diss,loc} 
\gtrsim 1$ and $\ell_{\rm diss,loc} > 0.04$ for $e^\pm$ and $ep$ plasmas, 
respectively. For lower values of $\ell_{\rm diss,loc}$, the dependence 
rolls over to approach the classical heat flux case in the limit of
very small compactness (in numerical calculations we obtain the classical
heat flux limit by setting $\phi_s=1000$).  Figure 3 shows the dependence of
$\epsilon_q$ on $\tau_{\rm tot}$. For lower local compactness parameters,
$\epsilon_q$ decreases with increasing $\tau_{\rm tot}$ due to a higher
radiative cooling efficiency of the corona.  For higher $\ell_{\rm diss,loc}$
(in the saturation regime), $\epsilon_q$ is only weakly dependent on $\tau_{\rm
tot}$; an increase of the cooling efficiency with $\tau_{\rm tot}$ is
compensated by an increase of the saturated heat flux (due to the increase of
$\tau_x$).
  
Solid curves on Figures 4 and 5 show the temperature, the dimensionless
electron density and the heat flux profiles in the corona ($x>0$) and in the
transition layer ($x<0$) obtained within our thermal conduction model.  The
dashed curves show the solutions of the model in which conduction is neglected
(the temperature here is determined by the equilibrium between heating and
cooling).

Saturation prevents the temperature in the corona from being much lower than
the Compton equilibrium temperature.  Namely, a decrease of the temperature
results in a decrease of the radiative cooling efficiency, which increases the
amount of energy that must be taken out of the corona by the conductive flux.
This quickly exceeds the capability of the saturated transport.  The amount by
which the corona temperature may deviate from the Compton equilibrium value
depends on the local compactness. At values of $\ell_{\rm diss,loc}$ for which
conduction is important, the difference may become significant (Figures 4 and
5, left panels). In this case, conduction leads to an observationally
interesting effect of the softening of the spectrum, as a lower plasma
temperature is achieved while the optical depth remains unchanged.  For higher
compactness parameters, only a tiny difference is possible and then most of
the coronal emission comes from a roughly isothermal region (Figures 4 and 5,
right panels).

The transition layer formed below the base of the corona has a relatively high
temperature and may extend over a height of a few tenths of the corona height
scale, as shown by solid curves at $x<0$ on Figures 4 and 5.  The existence of
such a layer was overlooked in previous studies of the disk-corona interface
(e.g., Shimura et al. 1995) as they neglected conduction, which appears to
provide the major contribution of energy to that region. Indeed, by
comparing the energy input to the transition layer due to the maximum
(saturated) value of conductive flux with the energy input due to heat flux of
coronal radiation, we find that
\begin{equation}
{q_{\rm sat} \over H_{\rm Compt}h_{\rm tl}}={5 \phi_s \delta \tau \Theta^{3/2}
\over 
\ell_{\rm h,loc} \langle \varepsilon \rangle \tau_{\rm tl} \eta} \left ( 
{m_e \over \mu} \right )^{1/2},
\end{equation}
where $\tau_{\rm tl}$ is the optical depth of the transition layer and
$\langle \varepsilon \rangle$ is given by eq.\ (\ref{vareps}). From this we
find that, e.g., for the parameters of the model shown on the right panel of
Figure 4 conduction provides 10 times more energy than Compton heating.  The
weak contribution of Compton heating is due to the fact that the average
energy of photons emitted by unsaturated Comptonization is much lower than the
average electron energy in that plasma. Then, if a purely radiative energy
exchange between two phases of a plasma occurs, the irradiated phase achieves
a much lower temperature than the hot one.  For a disk-corona system this would
lead to a sharp decline in the temperature profile, as shown by  dashed curves
on Figures 4 and 5.  The optical depth and Compton parameter, $y_{\rm
tl}=\int_{x_{\rm tl}}^0 (4\Theta+16\Theta^2)\tau_x dx$, of the transition
layer decrease with increasing local compactness (because a lower $y_{\rm tl}$
is needed at higher luminosity for the loss of the transported energy), e.g.
$\tau_{\rm tl} \gtrsim 0.1$ and $y_{\rm tl} \gtrsim 0.3$ for parameters
yielding $\epsilon_q > 0.2$, while $\tau_{\rm tl} < 0.01$ and $y_{\rm tl} <
0.05$ for $\epsilon_q < 0.05$.

A few solutions corresponding to different $\Theta_{\rm top}$ are shown on the
left panel of Figure 5. The solutions belonging to the same family may have
$y_{\rm tl}$  differing by a few percent depending on the relative importance
of Comptonization and bremsstrahlung in the loss of transported energy (see
Appendix A). Each of these solutions has slightly different pressure at the
base of the transition layer, which corresponds to a dimensionless ionization
parameter
\begin{equation}
\Xi= {F_h \over c P_{\rm gas}}.
\end{equation}
The actual solution of the corona structure is determined by the disk
solution, in particular by the resulting value of the ionization parameter at
the top of the disk.
The value of the ionization parameter at the base of transition layer is
relatively sensitive to the value of $\Theta_{\rm top}$, as e.g.\ a relative
change of the top temperature, $\Delta \Theta /\Theta \sim 10^{-4}$, results
in a change of ionization parameter, $\Delta \Xi / \Xi \sim 10^{-2}$, while
the transported power varies then by only $\Delta \epsilon_q/\epsilon_q
\sim 10^{-4}$.

As discussed in section 2.4, the value of $\phi_s$ is not exactly determined.
Therefore, we checked the sensitivity of the results to the value of $\phi_s$
(see Figure 6). It appears that the level of that sensitivity is determined by
the optical depth of the corona. For higher optical depths ($\tau \sim 0.3$),
the solution depends very weakly on the exact value of $\phi_s$. However, for
low optical depth ($\tau < 0.1$), $\epsilon_q$ increases significantly with
increasing $\phi_s$ for $\phi_s<0.6$, but levels off at higher
$\phi_s$.

\section{DISCUSSION AND SUMMARY}
We have developed a model of thermal conduction in a hot, active region
situated above the surface of much colder matter. The model is applicable to
regions with approximately plane parallel geometry so that the heat conduction
equation may be reduced to one dimension. It applies also to localized active
regions with, e.g., pillbox geometry (Stern et al. 1995), except for eq.\
(\ref{lhs}) in which the deficit of soft photons must be taken into account.

The importance of thermal conduction in accretion disk coronae depends
strongly on the local compactness parameter. For $ep$ plasma the conduction
has a negligible effect for $\ell_{\rm diss,loc} > 0.1$ independent of other
parameters. Pairs increase the importance of conduction by raising the sound
speed, and therefore the maximum (saturated) heat flux.  Our approach to pair
coronae is highly simplified (a pure $e^\pm$ corona and an $e^\pm$ transition
layer, no energy losses in a coronal wind) and is aimed to estimate the
maximum effect of thermal conduction. Even in this highly idealized situation,
conduction is important only for $\ell_{\rm diss,loc}<1$.
 
The value of $\ell_{\rm diss,loc}$, 
crucial for the conduction problem, depends on
the luminosity and geometrical parameters of the active region. E.g., for an
electron-proton corona extending over the total surface of the accretion disk
($A \simeq \pi R^2$) we find from the definition of the local compactness,
using eq.\ (\ref{hr}) and $r \sim 20$,
\begin{equation}
\ell_{\rm diss,loc}=54 \Theta_c^{1/2} {L \over L_{\rm Edd}}.
\end{equation}
Then for $10^{-3}<L/L_{\rm Edd}<1$ and $0.2<\Theta <1$ we can expect local
compactnesses ranging from 0.02 to 54. In this case conduction is important
for $L \lesssim 10^{-2} L_{\rm Edd}$, a condition which may apply in at least
some AGN.  On the other hand, a luminosity of $\sim 10^{-2} 
L_{\rm Edd}$ is typical, e.g., for the hard state of the Galactic
black hole candidate Cyg X-1 (e.g., Phlips et al.\ 1996).
Thus, conduction is likely to be important in at least some AGNs and Galactic
black hole candidates.

When conduction becomes important, a nonuniform temperature distribution is
established in the corona. In addition, a relatively hot transition layer
between the corona and the disk is formed. In this case, the coronal radiation
has a spectrum steeper than in models without conduction (with the same
$\tau_{\rm tot}$ and $\ell_{\rm diss,loc}$). Moreover, the spectrum will have a
more complicated form than a simple power-law. In particular, we expect a soft
power-law-like emission from the transition layer due to Comptonization of soft
photons from the disk in addition to the coronal emission.  The spectral index
of this emission should be larger than the spectral index of the coronal
emission, as the Compton parameter of the transition layer is lower than the
Compton parameter of the corona, except for very low local compactness
$\ell_{\rm diss,loc} \lesssim 0.01$. The superposition of the transition layer
component on the spectrum of the coronal radiation (the latter extends to
higher energies due to higher temperature in the corona) is a possible
explanation of the formation of soft X-ray excesses, observed in many AGNs at
energies below $\sim 1$ keV (e.g., Wilkes \& Elvis 1987; Turner \& Pounds
1989). When conduction is not important, the corona is isothermal and a rather
sharp transition occurs at the corona base to the region in radiative
equilibrium.

\acknowledgements
This research has been supported by Polish KBN grants 2P03D01008
and 2P03D01111 and NASA grant NAGW-3156.

\appendix

\section{$\Theta_{\rm top}$}
We discuss here the criteria for adjusting $\Theta_{\rm top}$.

$\Theta_{\rm top}$ determines the fraction of energy released in the corona
that must be transported out of corona by conduction.  For a lower $\Theta_{\rm
top}$, Comptonization is less efficient and a higher flux must be carried down
at lower plasma conductivity ($\propto \Theta^{5/2}$), which requires higher
temperature gradient.  Thus the decrease of $\Theta_{\rm top}$ implies both the
increase of the heat flux at the corona bottom, $q_0$, and the decrease of the
temperature at the corona bottom, $\Theta_0$. The total energy transported by
conductive flux is radiated away in the transition layer.  At the base of the
transition layer the solution should have a vanishing conductive flux (as no
flux can be transported further down due to very small plasma conductivity) and
small temperature (determined by radiative equilibrium).  Below we present
analytic solutions for the transition layer structure in certain approximations
to show qualitative properties of the solution, which constrain the possible
value of $\Theta_{\rm top}$.

We consider the classical heat flux limit approximation with constant pressure,
$P_{\rm gas}=$const, and we neglect Compton heating.  Below we use the
dimensionless heat flux and pressure,
\begin{equation}
\tilde{q} \equiv -q_{\rm class}h{\sigma_T \over m_e c^3}=a\Theta^{5/2} \Theta',
\label{a1}
\end{equation}
\begin{equation}
\tilde{P} \equiv P_{\rm gas} h {\sigma_T \over m_e c^2} = 2\Theta \tau_x.
\end{equation}
For most of the transition layer, the temperature is sufficiently high (and
the electron density sufficiently low) for cooling to be completly dominated
by Comptonization. We consider then the heat transfer equation with the
cooling term determined by Compton cooling,
\begin{equation}
{d\tilde{q} \over dx} = 4 \Theta \tau_x \ell_{\rm diss,loc}(\ell_s+\ell_h)/
\ell_{\rm diss}.
\label{a2}
\end{equation}
Eqs.\ (\ref{a1}) and (\ref{a2}), with the boundary conditions determined by the
solution at the corona bottom, $\tilde{q}_0$ and $\Theta_0$, have the following
solution
\begin{equation}
\Theta=\left ( {7\ell \tilde{P} \over 2a}x^2 + {7 \tilde{q}_0 \over 2a}x + 
\Theta_0^{7/2} \right )^{2/7},~~~~~~~~\tilde{q}=2 \ell \tilde{P}x+\tilde{q}_0, 
\end{equation}
where $\ell \equiv \ell_{\rm diss,loc}(\ell_s+\ell_h)/\ell_{\rm diss}$.  We see
that for a given value of $\tilde{q}_0$ there is a unique value of
$\Theta_0^{\rm max}=[7\tilde{q}_0^2/(8a\ell \tilde{P})]^{2/7}$ for which
vanishing of $\tilde{q}$ at the bottom of the transition layer is accompanied
by $\Theta$ decreasing  to a very small value.  For higher $\Theta_0$
temperature does not decrease and if we tried to prolong the solution further
down the temperature would increase implying a negative temperature gradient.
The physical meaning of that class of solutions [with $\Theta_0 > \Theta_0^{\rm
max}(\tilde{q}_0)$] is that conductive transport from the disk is required, in
addition to $\tilde{q}_0$, in order to provide sufficient amount of energy to
the transition layer to balance the Compton cooling rate determined by
$\Theta_0$.  On the other hand, for $\Theta_0<\Theta_0^{\rm max}$ the
temperature vanishes before the transported thermal energy can be radiated away
due to Comptonization (in this case, the decrease of $\Theta$ to zero is
accompanied with divergence of $\Theta'$ fast enough to keep the heat flux at a
finite value). The decrease of $\Theta$ implies, however, the increase of
$\tau_x$ and thus bremsstrahlung becomes efficient as a cooling mechanism. To
check whether bremsstrahlung may provide sufficiently high cooling rate for the
loss of the rest of the transported thermal energy we consider the heat
conduction equation with the cooling term determined by bremsstrahlung. We use
the bremsstrahlung cooling rate for the nonrelativistic case and neglect the
second term in parentheses in eq.\ (\ref{lb}), namely $1.781\Theta^{1.84}$, for
calculational simplicity.  Then,
\begin{equation}
{d\tilde{q} \over dx}= \ell_b \tau_x^2 \Theta^{1/2},
\label{a3}
\end{equation}
where $\ell_b = 4 \sqrt{2} \alpha_f \pi^{-3/2}$.
Eqs.\ (\ref{a1}) and (\ref{a3}) have the solution,
\begin{equation}
\tilde{q}=a \left[{\ell_b \tilde{P}^2 \over 4a} (\Theta^2 - \Theta_b^2) +
\left ( {\tilde{q}_b \over a} \right )^2 \right ] ^{1/2},
\end{equation}
where $\tilde{q}_b$ and $\Theta_b$ are the heat flux and the temperature at the
point where bremsstrahlung starts to dominate. For $\Theta_b< 2\tilde{q}_b/
\sqrt{\ell_b \tilde{P}^2 a}$, the energy transported by the conductive flux
cannot be radiated away. This means  that certain $\Theta_0^{\rm min}$ exists
such that for lower $\Theta_0$ no acceptable solution exists, because
$\tilde{q}_b$ and $\Theta_b$ are positively correlated with $\tilde{q}_0$ and
$\Theta_0$, respectively.

The solution of eqs.\ (\ref{e1}--\ref{e3}) has the same qualitative properties
as the approximation we investigated above. A certain range of the values of
$\Theta_0$ exists then, $(\Theta_0^{\rm min}, \Theta_0^{\rm max})$, for which
we obtain physically acceptable solutions. For lower $\Theta_0$, the energy
that enters the transition layer cannot be radiated away and as a result the
temperature in the layer would increase leading to the increase of $\Theta_0$.
On the other hand, for higher $\Theta_0$ the energy delivered to the layer is
too low to maintain its temperature and the temperature would decrease.
Solutions corresponding to different $\Theta_0^{\rm min}< \Theta_0<
\Theta_0^{\rm max}$ differ with respect to the relative importance of
bremsstrahlung and Compton cooling in the loss of transported energy.

As discussed above, the decrease of $\Theta_{\rm top}$ implies a decrease of
$\Theta_0$ and an increase of $\tilde{q}_0$. As the value of $\tilde{q}_0$ is
constrained by the saturation effect (see section 4), the above constraints on
$\Theta_0$ imply the existence of a certain range of the values of $\Theta_{\rm
top}$ for which the correct behaviour of the solution in the transition layer
can be obtained. In our calculation procedure (section 2.7), we determine this
range by looking at the behaviour of the solution at the bottom end for
different values of $\Theta_{\rm top}$.

\newpage

\figcaption{The dependence of $\epsilon_q$ on $\ell_{\rm diss,loc}$ in
electron-positron plasmas for the value of $\phi_s=0.3$ (solid curves)
compared to the classical heat flux approximation (dotted curve). For solid
curves $\tau_{\rm tot}=0.06$, 0.1 and 0.2 from top to bottom and for the
dotted curve $\tau_{\rm tot}=0.1$. For $\ell_{\rm diss,loc} > 1$, $\epsilon_q
\propto \ell_{\rm diss,loc}^{-1}$ for the solid curves.
\label{fig1}} 

\figcaption{The same as in Figure 1 but for electron-proton
plasmas with $\tau_{\rm tot}=0.1$; $\epsilon_q \propto \ell_{\rm
diss,loc}^{-1}$ for $\ell_{\rm diss,loc} > 0.04$.
\label{fig2}}

\figcaption{$\epsilon_q$ as a function of the optical depth of the corona.
\label{fig3}}

\figcaption{The corona and transition layer structure obtained in our model 
(solid curves) and the solution of the model neglecting conduction (dotted
curves) for electron-proton coronae with $\ell_{\rm diss,loc}=0.01$ (left
panel) and $0.03$ (right panel) and $\tau_{\rm tot}=0.1$.  The curves show (a)
the temperature, (b) the electron density in the units of $\tau_x= n_e(x)
\sigma_T h$, and (c) the heat flux absolute value $|q|$ with respect to the
power dissipated per surface area, $L_{\rm diss}/A$. For $\ell_{\rm
diss,loc}=0.01$ solution is obtained for $\Theta_{\rm top}=0.52466$. 
The solution gives
$\epsilon_q=0.5$,  the ionization parameter at the base of the transition
layer, $\Xi=0.004$, and the optical depth and Compton parameter in the
transition layer, $\tau_{\rm tl}=0.4$ and $y_{\rm tl}=0.6$, respectively.  The
absolute range of possible $\Theta_{\rm top}$ is $\Delta \Theta = 1.5
\times 10^{-5}$, which yields $\Delta \epsilon_q=2 \times
10^{-5}$ and $\Delta \Xi = 1.5 \times 10^{-4}$. 
For $\ell_{\rm loc}=0.03$, the solution, with $\Theta_{\rm top}=0.618353$,
gives $\epsilon_q=0.2$,  $\Xi=0.04$, $\tau_{\rm tl}=0.2$, $y_{\rm tl}=0.3$,
$\Delta \Theta = 1 \times 10^{-6}$, $\Delta \epsilon_q=3
\times 10^{-6}$, $\Delta \Xi = 10^{-3}$. 
The break of $|q|A/L_{\rm diss}$ at the interface between the transition layer
and the corona ($x=0$) is due to the fact that the corona is efficiently
heated by dissipation, whereas only the much smaller Compton heating takes
place in the transition layer.  
\label{fig4}}

\figcaption{The same as in Figure 4 but for electron-positron plasmas,
$\ell_{\rm diss,loc}=0.1$ (left panel) and $1$ (right panel) and $\tau=0.1$.
For $\ell_{\rm diss,loc}=0.1$, the solution gives $\Theta_{\rm top}=0.63155$,
$\epsilon_q=0.2$,
$\Xi=0.3$, $\tau_{\rm tl}=0.12$ and $y_{\rm tl}=0.3$,  $\Delta \Theta = 5
\times 10^{-5}$, $\Delta \epsilon_q=3 \times 10^{-4}$,
$\Delta \Xi = 10^{-2}$. For $\ell_{\rm diss,loc}=1$, we get
$\Theta_{\rm top}=0.650531$, $\epsilon_q=0.05$,  $\Xi=3.6$,
$\tau_{\rm tl}=0.02$ and $y_{\rm tl}=0.05$,  
$\Delta \Theta = 10^{-6}$, $\Delta \epsilon_q=3 \times 10^{-4}$,
$\Delta \Xi = 1.5 \times 10^{-2}$.
The three solid curves on the left panel show
solutions corresponding to three different values of
$\Theta_{\rm top}$. 
\label{fig5}}

\figcaption{The dependence of $\epsilon_q$ on $\phi_s$. Left panel:
electron-positron plasmas, $\ell_{\rm diss,loc}=0.12$, 0.12, 1.2, 1.2 and
$\tau_{\rm tot}=0.06$, 0.3, 0.06, 0.3  from top to bottom. Right-panel:
electron-proton plasmas, $\ell_{\rm diss,loc}=$0.01, 0.01, 0.08, 0.08 and
$\tau_{\rm tot}=0.1$, 0.32, 0.1, 0.32 from top to bottom.
\label{fig6}}   

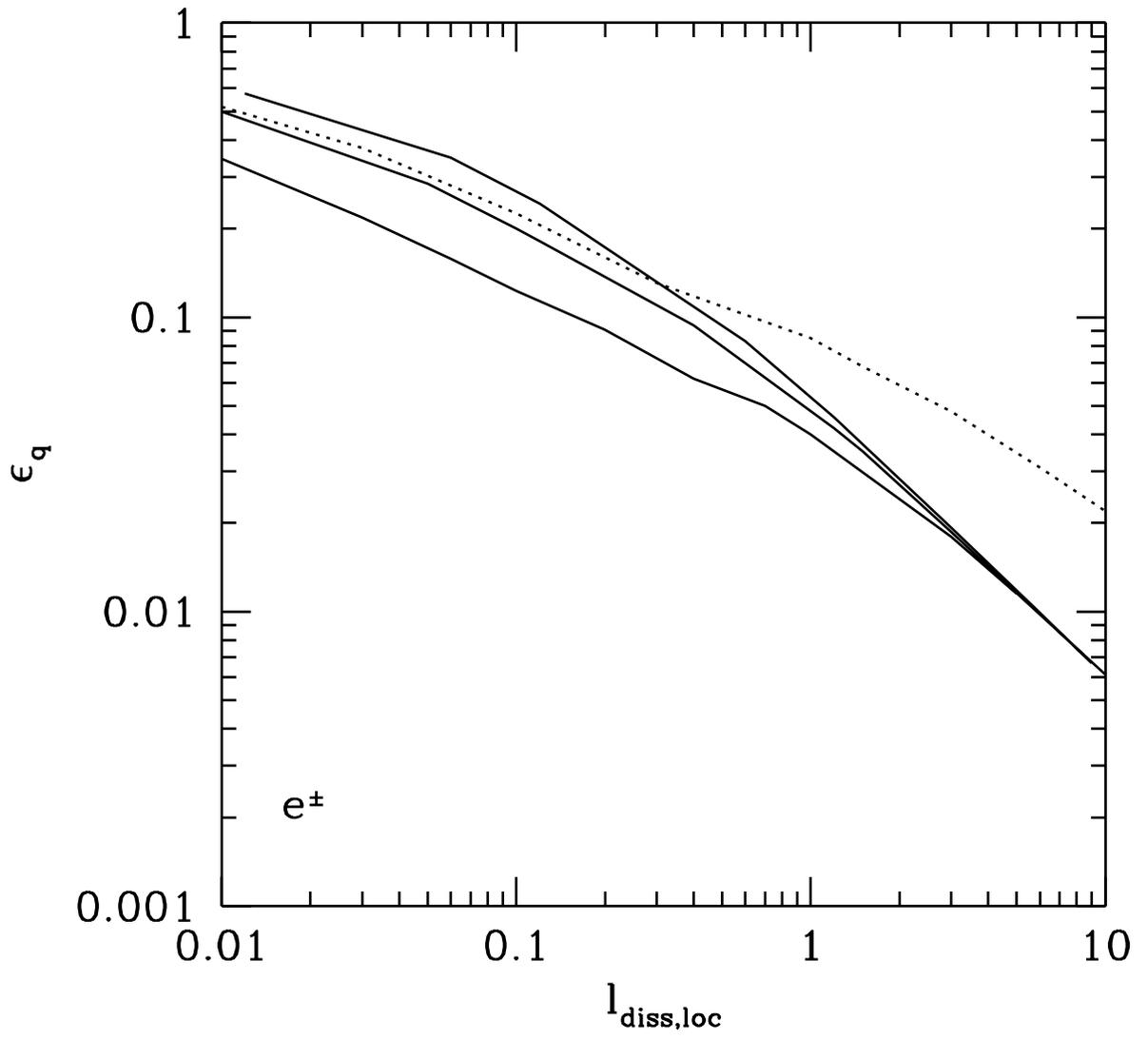
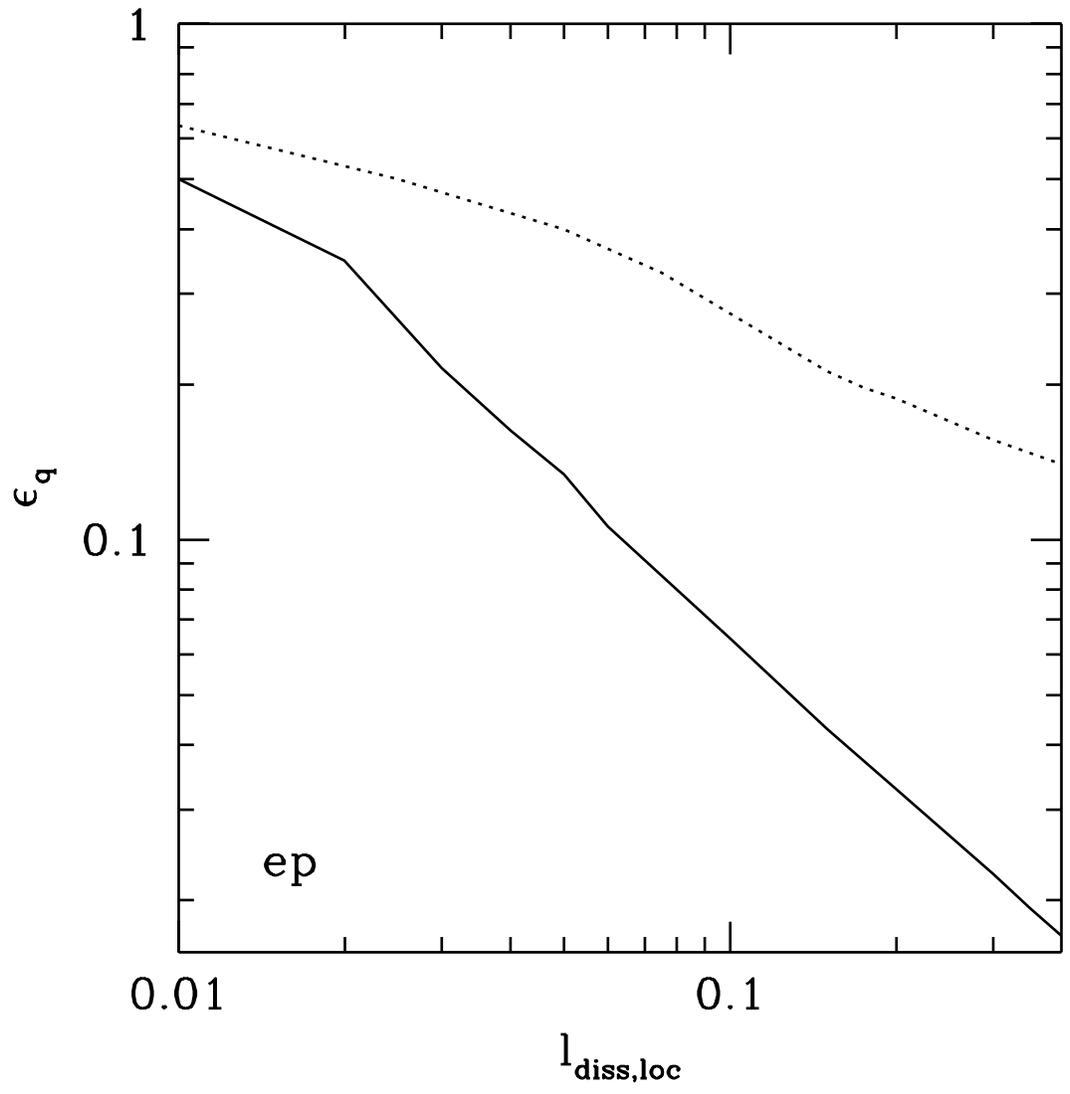
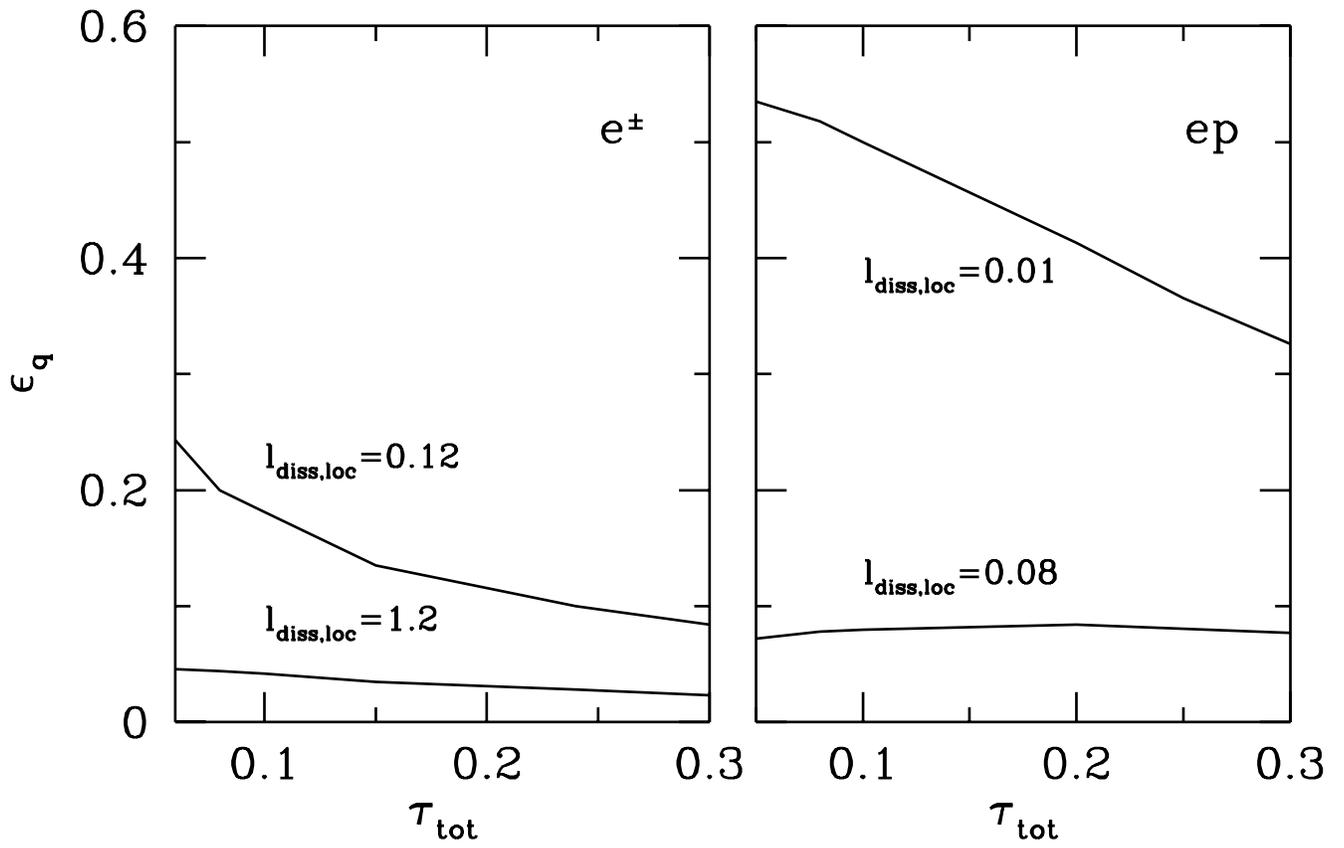
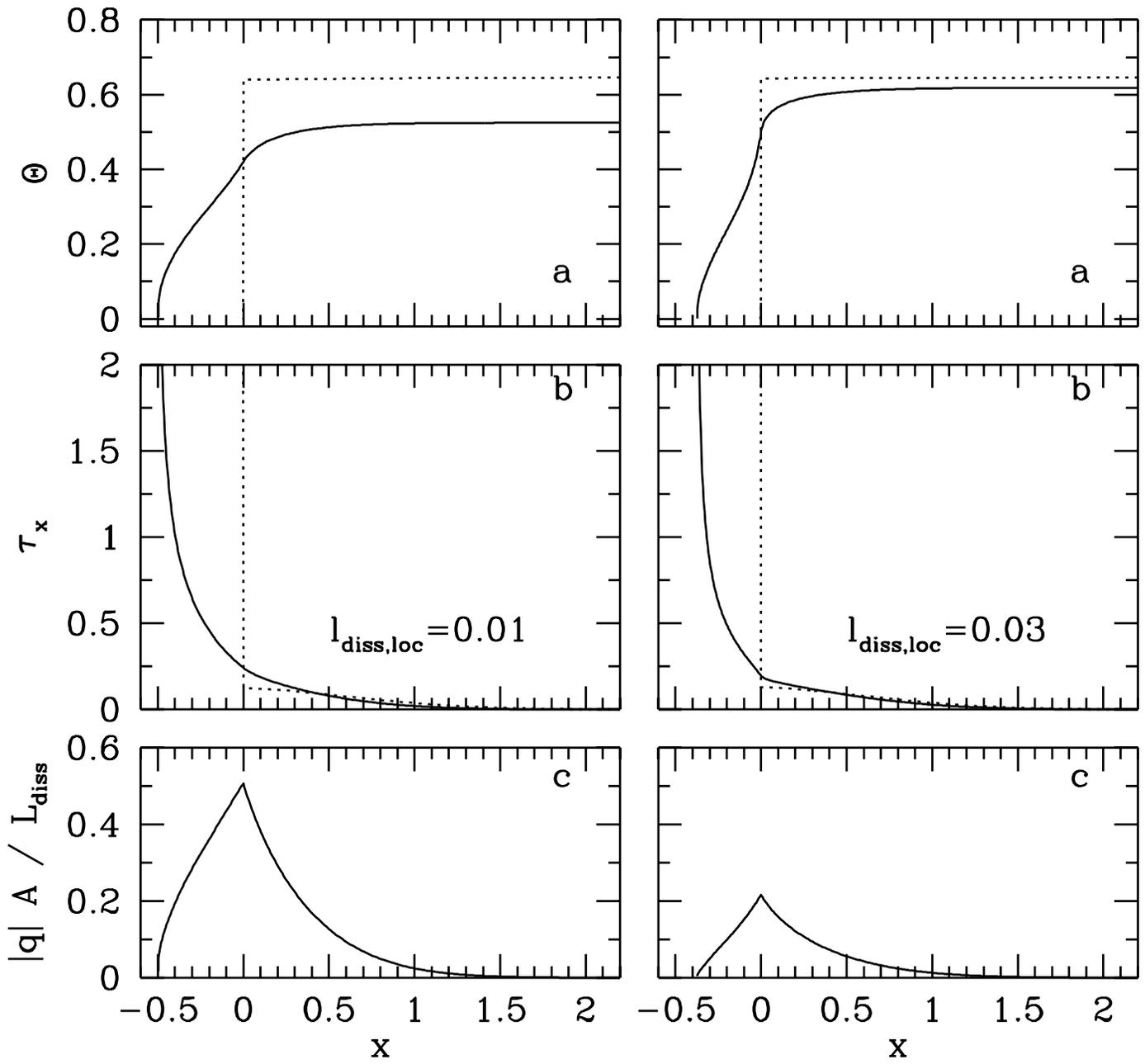
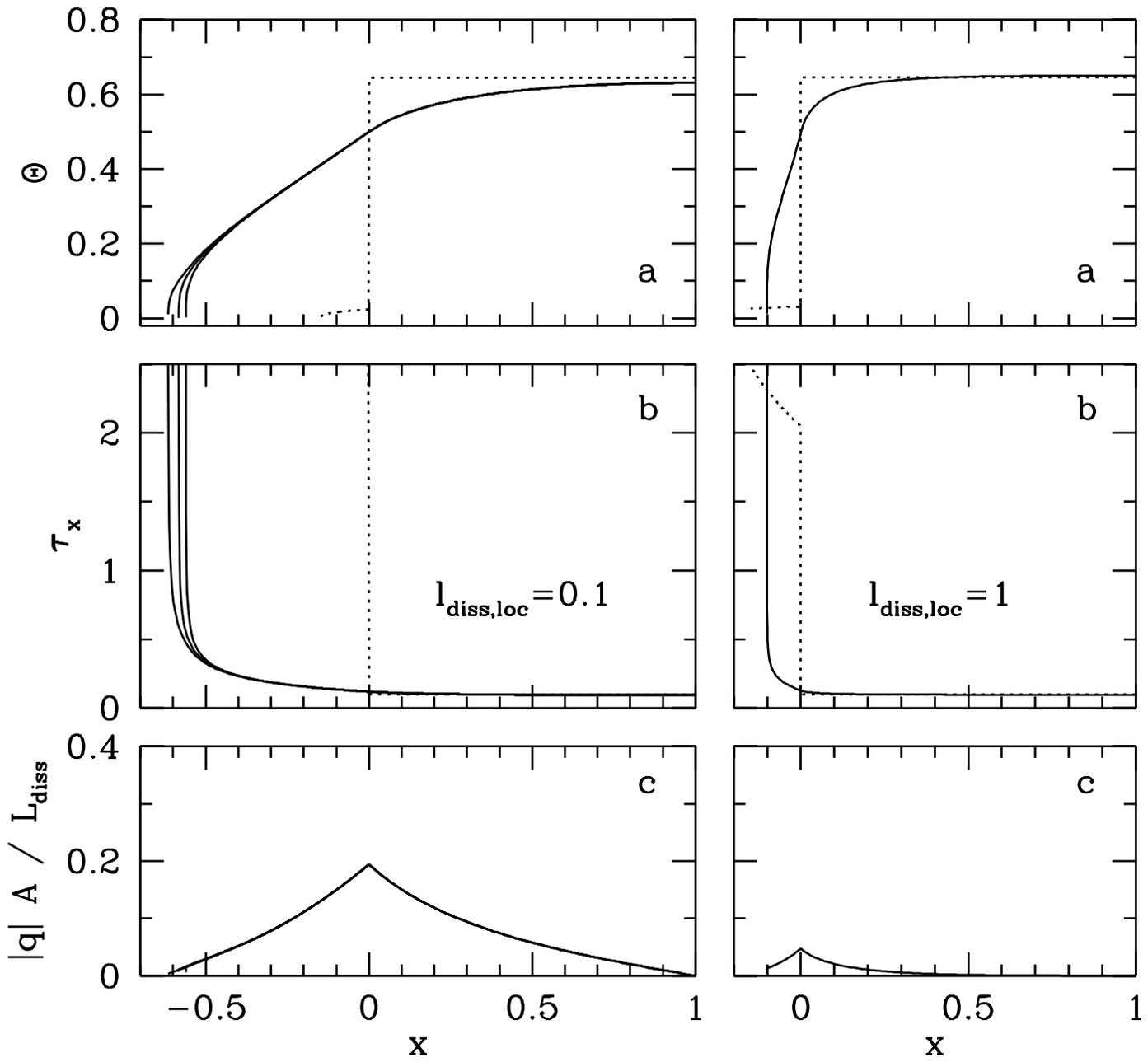
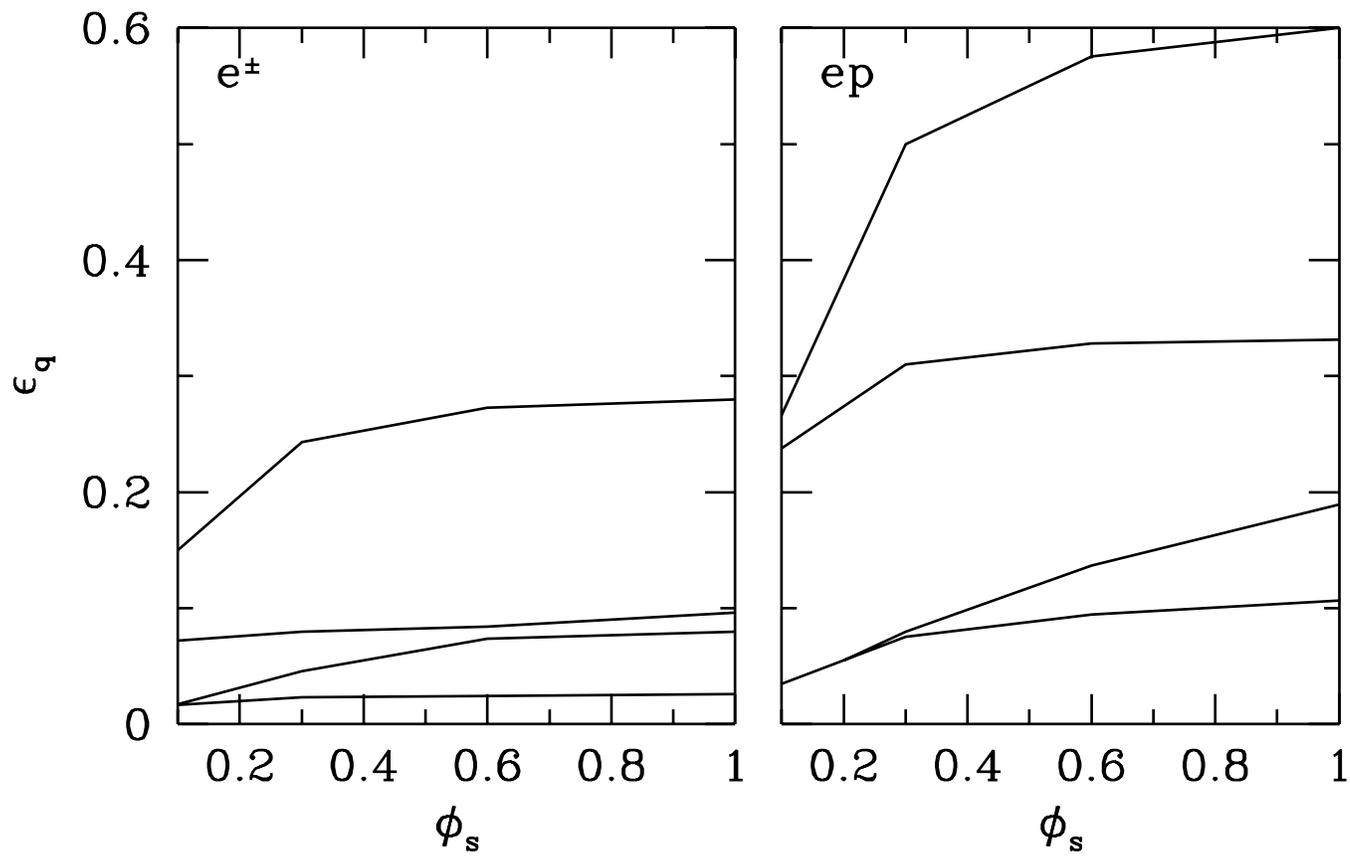
\end{document}